\documentclass[doublecol,A4paper]{epl2}

\title{Unconventional temperature dependence of the cuprate \\excitation spectrum}

\author{William Sacks\inst{1} \and Alain Mauger\inst{1} \and Yves Noat\inst{2}}
\shortauthor{W. Sacks, A. Mauger \& Y. Noat}

\institute{ \inst{1} Institut de Min\'{e}ralogie, de Physique des
Mat\'{e}riaux, et de Cosmochimie (IMPMC), UMR 7590, Sorbonne
Universit\'{e}s, UPMC Paris 6, 4 place Jussieu, 75252 Paris Cedex 05,
France.

\inst{2} Institut des Nanosciences de Paris (INSP), UMR 7588,
Sorbonne Universit\'{e}s, UPMC Paris 6, 4 place Jussieu, 75252 Paris
Cedex 05, France. }

\abstract{Key properties of the cuprates, such as the pseudogap
observed above the critical temperature $T_c$, remain highly
debated. Given their importance, we recently proposed a novel
mechanism based on the Bose-like condensation of mutually
interacting Cooper pairs [W. Sacks, A. Mauger, Y. Noat,
Superconduct. Sci. Technol. 28 105014, (2015)]. In this work, we
calculate the temperature dependent DOS using this model for
different doping levels from underdoped to overdoped. In all
situations, due to the presence of excited pairs, a pseudogap is
found above $T_c$ while the normal DOS is recovered at $T^*$, the
pair formation temperature. A similar behavior is found as a
function of magnetic field, crossing a vortex, where a pseudogap
exists in the vortex core. We show that the precise DOS shape
depends on combined pair (boson) and quasiparticle (fermion)
excitations, allowing for a deeper understanding of the SC to the PG
transition.}


\begin{document}

\maketitle

\section{Introduction}

As is well known, the superconducting (SC) state of cuprates is
characterized by a dome-shaped critical temperature $T_c$ versus
carrier density and an unconventional pseudogap (PG) state above
$T_c$ (see \cite{Timusk_rev} for a review). In addition, these
compounds exhibit a spatially inhomogeneous
\cite{davis_rev,Sci_McElroy2005} and unconventional quasiparticle
(QP) dispersion (see
\cite{RevMod_Damascelli2003,Natphys_Hashimoto2014,fischer_rev} and
Ref. therein), the `peak-dip-hump' structure, with a characteristic
gap $\Delta_p$ typically larger than the critical temperature energy
scale $\sim k_B\,T_c$, especially in the underdoped regime.

Many of these properties remain highly debated, in particular the
Cooper pair formation mechanism and the origin of the pseudogap
state, which appears below the higher temperature $T^*$. The
spectral gap is of the same order of magnitude at $T_c$ as in the
vortex core \cite{PRL_renner1998_B}, indicating that the PG
characterizes the loss of SC coherence \cite{PRL_Miyakawa1999}. In
agreement with Renner et al.\cite{PRL_renner1998_T}, recent
experiments by Sekine et al. \cite {JPSJ_Sekine2016} on
quasiparticle excitation spectra up to $T^*$ are particularly
intringuing. The spectral gap indeed starts to disappear and
vanishes only at the higher temperature $T^*$.

To summarize, a solution of the high-$T_c$ problem must account for\,: \parindent 12 pt

\vspace{0,2cm}

($i$) multiple energy scales ($T_c$, $T^*$) and large gap $\Delta_p$,

($ii$) large $T_c$, however with $k_B\,T_c \ll \Delta_p$,

($iii$) the dome-shape of the $T-p$ phase diagram,

($iv$) the unconventional SC to PG transition,

($v$) the unusual QP dispersion ( `peak-dip-hump').

\parindent 12 pt
\vspace{0,2cm} With these questions in mind, we have recently
proposed a model \cite{physicaC_Sacks2014,SciTech_Sacks2015}, based
on the mutual interaction between pre-formed pairs, able to take
into account both phases of the system. It describes cogently the
shape of the low-temperature excitation spectrum in cuprates and
links the superconducting and pseudogap phases.

In an early approach, Franz and Millis
\cite{PRB_Franz1998} addressed the temperature dependence of the QP
excitation spectrum based on the effects of classical phase
fluctuations. Unfortunately, their model does not match the precise
shape of the STS data, especially the important dip present at low
temperature, a clear signature of HTc superconductivity.

More recently, Pieri et al. \cite{PRB_Pieri2004}
considered pair fluctuations in the t-matrix framework, but again no
dip in the zero temperature spectra was clearly demonstrated. In a
related paper involving pair fluctuations \cite{PhysicaB_Letz1997},
a repulsive pair-pair interaction was inferred, like in our model,
but an explicit calculation of the density of states was not
reported. To the contrary, our model gives for the first time the
low-temperature peak-dip-hump feature and the full temperature
dependence of the QP spectra up to $T^*$.

\begin{figure}[t]
\centering
\includegraphics[width=8.cm]{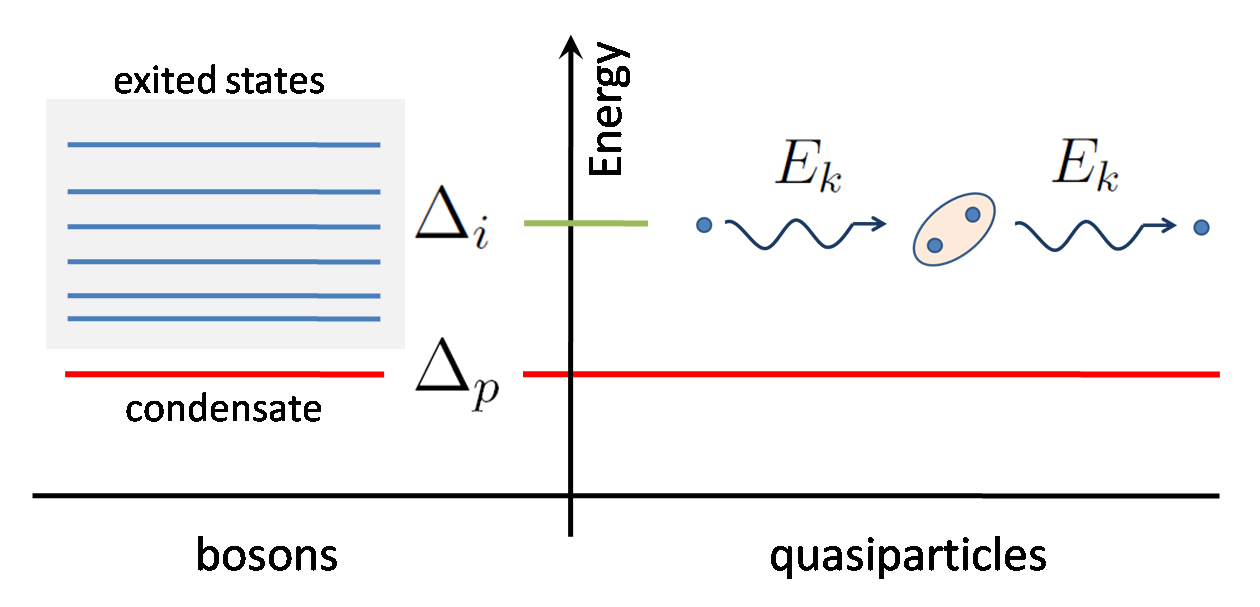} \caption{Left panel\,: distribution of
excited pair states. Right panel\,: energy diagram to illustrate the
coupling between quasiparticles and excited pair states. Such a
strong QP-pair coupling gives rise to the second term in the gap
equation\,\ref{gap_equa}.}\label{Fig1}
\end{figure}

At the heart of the model, an incoherent state consists of preformed
pairs \cite{Nat_Emery1995} with an energy distribution
$P_0(\Delta_i)$ (see Fig.\,\ref{Fig1}, left panel):
\begin{equation}
P_0(\Delta_i) \propto \frac{\sigma_0^2}{(\Delta_i-\Delta_0)^2 +
\sigma_0^2}
\end{equation}\label{P0}
where $\Delta_0$ and $\sigma_0$ are respectively the average gap and
the width of the distribution. Alternatively, $P_0(\Delta_i)$ can be
considered as the density of pair states.

Pairs in the PG state have a random energy distribution, making long
range order impossible. Interactions between pairs, with
characteristic coupling energy $\beta_k$, allows them to couple and
condense. The superconducting state is achieved through a Bose-type
condensation in a single quantum state with a well-defined global
phase, where the superconducting gap $\Delta_k$ is determined by the
self-consistent equation \cite{SciTech_Sacks2015}:
\begin{equation}
\Delta_k(E_k)=\Delta_{0,k}-2\,\beta_k P_0(E_k)
\label{gap_equa}
\end{equation}
with $E_k=\sqrt{\epsilon_k^2+\Delta_k^2}$, the quasiparticle dispersion.
The self-consistent solution of this equation at $\epsilon_k=0$ gives the condensation level $\Delta_k=\Delta_p\cos(2\theta)$. In the antinodal direction ($\theta=0$), the condensation energy is thus $$\varepsilon_c=\Delta_0-\Delta_p=2\beta P_0(\Delta_p)$$
The unconventional dependence of the interaction term $\beta_k P_0(E_k)$ with the quasiparticle energy $E_k$ expresses the interaction of a quasiparticle with pair excited states (see the diagram, Fig.\,\ref{Fig1}).

The coefficient $\beta_k\propto\beta_0 N_{oc}(T)$ is proportional to
the number of pairs in the condensate $N_{oc}(T)$ and plays the role
of an order parameter vanishing at $T_c$. Taking the {\it d}-wave
symmetry of the SC wave function into account, $N_{oc}(T)$ is given
by \cite{SciTech_Sacks2015}:
\begin{eqnarray}
&N_{oc}(T)& =n_0-\mathcal{A}(T) \,\int_{0}^{2\,\pi}d\,\theta
\int_{\Delta_p
{\rm cos}(2\theta)+\delta}^{\infty}\,{\rm d}\Delta_i\ \hskip 0.1 cm \null \nonumber\\
&\null& \times \,  P_0(\Delta_i) f_B(\Delta_i-\Delta_p\,{\rm
cos}(2\theta), T)\label{Noc}
\end{eqnarray}
where $f_B(E,T)=(e^\frac{(E-\mu)}{k_BT}-1)^{-1}$ is the Bose
statistics, $\Delta_p$ the ground state energy and $\mathcal{A}(T)$
is for normalization. The `minigap' $\delta$ separates the
condensate from the first excited state and is thus identified as
the energy needed to extract a pair from the condensate. As we shall
examine below, this quasi-Bose transition is quite distinct from the
atomic situation, due to coexisting quasiparticle and normal
electron excitations of the condensate.

The interaction term $\beta_k P_0(E_k)$ in the gap
equation\,\ref{gap_equa} is responsible for the `peak-dip-hump'
structure in the QP excitation spectrum. The fit of the QP DOS using
equation\,\ref{gap_equa} allows to extract the SC parameters
($\Delta_0$, $\sigma_0$, $\Delta_p$, $\beta$) with high precision as
a function of doping, i.e. the phase diagram. As shown in
\cite{SciTech_Sacks2015}, at $T=0$, coherence is not revealed in the
spectral gap $\Delta_p$ but in the fine structure (`peak-dip-hump')
developing at higher energy.

In this work, we extend the notion of quasiparticle DOS usually
associated with the condensate to include the excited states as
well. In particular, we calculate the temperature dependence of the
QP DOS from the low-$T$ condensed state through the PG state and up
to $T^*$, for a range of doping values from underdoped to overdoped
regimes. We show that the contribution of excited states gises rise
to a pseudogap at $T_c$, in full agreement with experiments
\cite{PRL_renner1998_T}. In addition, a similar result is found in
the vortex core, where coherence is broken by the magnetic field.
The coexistence of boson and fermion excitations and their relative
contribution to the DOS are the main points of the present work. Our
model thus reproduces the temperature dependence of the experimental
STM spectra up to $T^*$ (see
\cite{JPSJ_Sekine2016,PRL_renner1998_T}). It gives direct insight
into the new condensate density and the excited-pair density which
are original to the model.
\begin{figure*}[t]
\centering
\includegraphics[width=8.75 cm]{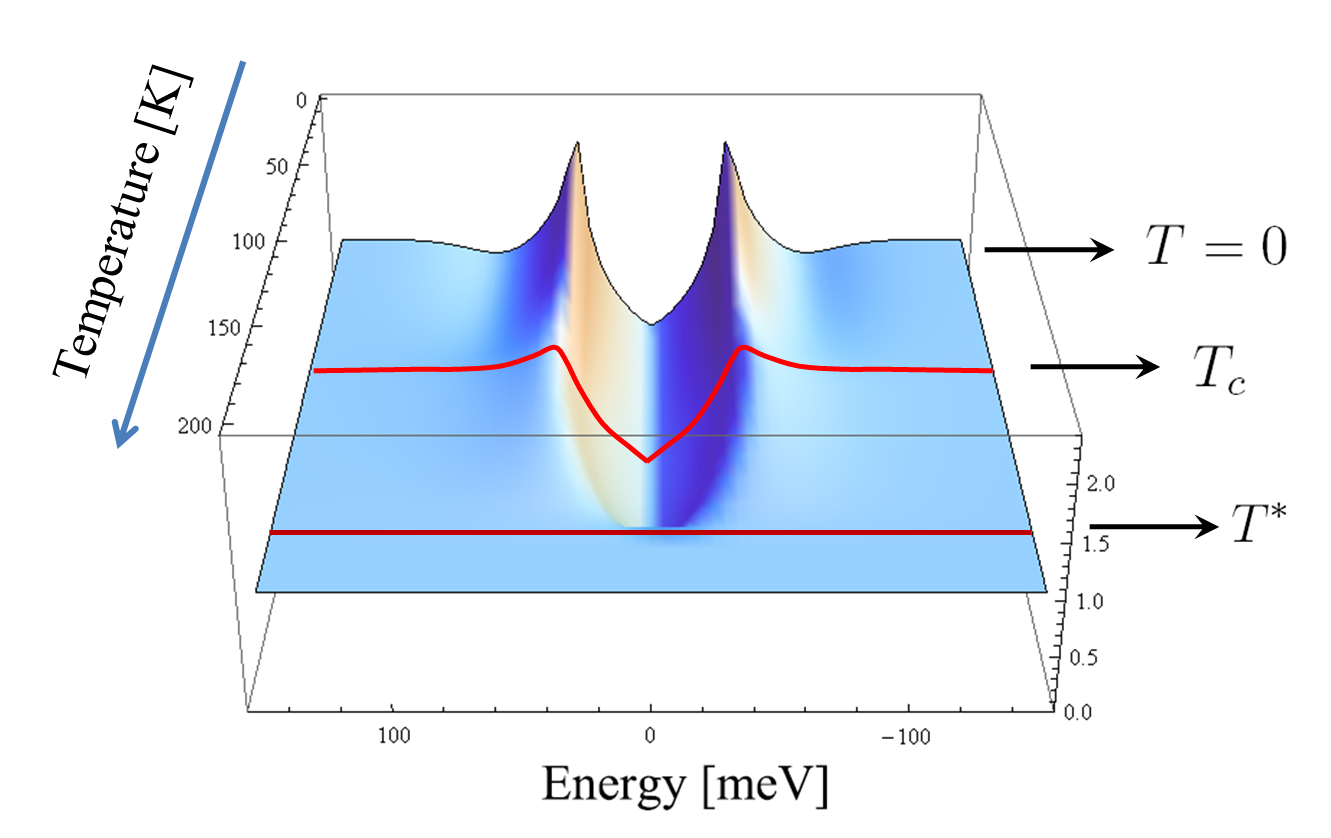}
\includegraphics[width=7. cm]{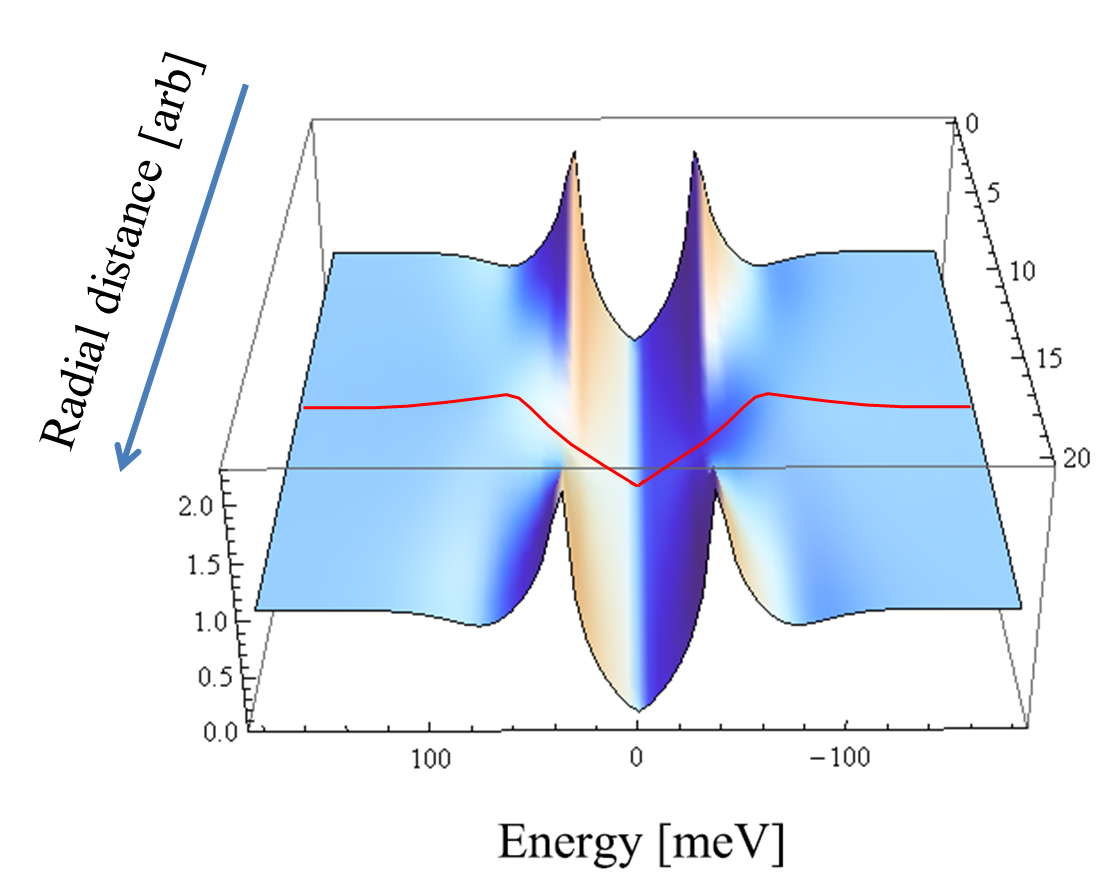}
\caption{Left panel\,: temperature evolution of the QP-DOS from zero
temperature through $T_c$ (SC to PG transition) up to $T^*$ (normal
state). Right panel\,: expected evolution of the QP-DOS crossing a
vortex core. The coherence peaks disappear leaving a peakless PG
spectrum. Note that the additional thermal broadening ($\sim
3.5\,k_BT_c$) due to the tunneling process is not included
throughout this work.}\label{Fig3}
\end{figure*}

\section{Quasiparticle excitation spectrum}

\subsection{Case of conventional superconductors}
The excitations of the system are important for all thermodynamic and transport properties. In the BCS theory \cite{BCS}, at finite temperature, the presence of condensed Cooper pairs below $T_c$ is accompanied by quasiparticle excitations having the dispersion $E_k=\sqrt{\epsilon_k^2+\Delta_k(T)^2}$ where $\epsilon_k$ is the kinetic energy relative to the Fermi energy and $\Delta(T)$ is the SC gap. The latter decreases as a function of temperature as a result of QP excitations governed by Fermi-Dirac statistics $f(E,T)=(e^\frac{(E-\mu)}{k_BT}+1)^{-1}$ and eventually vanishes at the critical temperature $T_c$.

The zero-temperature gap is related to the fundamental parameters and energy scales of the material (Debye frequency, electron-phonon coupling, DOS at the Fermi energy). Proportional to the critical temperature, it has a universal value : $\Delta(T=0)\approx 1.7\,k_B\,T_c$.

The tunneling density of states (DOS) is a direct consequence of the QP dispersion:

\begin{equation}
\mathcal{N_S}(E)=\mathcal{N}_n(E_F)\frac{|E|}{\sqrt{E^2-\Delta(T)^2}}\,\Theta(|E|-\Delta)
\end{equation}
where $\mathcal{N}_n(E_F)$ is the normal DOS at the Fermi energy and $\Theta(z)$ the heaviside function.
\\At low temperatures, it reveals a sharp energy gap at the Fermi-level - first measured by I. Giaever using planar junctions \cite{PRL_Giaever1960} - delimited by high symmetric peaks. The DOS has since been measured many times with a high precision by  tunneling spectroscopy using planar junctions or STM geometries \cite{APL_Pan1998} and later on using Angle Resolved Photon Emission Spectroscopy (ARPES) in cuprates \cite{PRL_Shen1993,PRL_Ding1995}.

\subsection{Unconventional superconductors}

In our model, the situation is radically different from the BCS situation. In this case, the gap does not vanish at $T_c$ but at $T^*$, the pair formation onset temperature. On the other hand, the condensation energy $\varepsilon_c\propto  \beta_0 N_{oc}(T)$ vanishes at $T_c$, which is much smaller than $T^*$. Thus, for the temperature dependent DOS, two additional contributions arise, absent in the BCS case: first the contribution of excited pairs increasing with temperature up to $T_c$, and second the dissociation of pairs increasing up to $T^*$.

The DOS as a function of temperature for optimal doping is reported in Fig.\,\ref{Fig3}, left panel. At zero temperature, it exhibits the characteristic shape of the excitation spectrum of the cuprates, with the characteristic `peak-dip' features. As the temperature rises, the quasiparticle peak heights decrease while the {\it gap width} remains roughly constant, contrary to the BCS case. At $T_c$, there is a clear pseudogap with attenuated QP peaks. For higher temperature ($T>T_c$), one notes the filling of states inside the gap while the QP peaks are further smoothed. The gap eventually disappears completely at $T^*$.

Note that other theories have been proposed for the `peak-dip-hump'
structure in the DOS measured by STM/ARPES. The most popular
approach is the coupling to a spin collective mode (see
\cite{PRL_Ahmadi2011}, \cite{AdvPhys_Eschrig2006} and reference
therein). It appears that the fit to the experimental spectra is
rendered difficult in this framework due to the necessarily retarded
interaction. Furthermore, to our knowledge, its extension to finite
temperature has not been done.

To the contrary, the pair-pair interaction model accurately
reproduces the DOS measured by STM/STS in
BiSr$_2$Ca$_2$Cu$_2$O$_{8+\delta}$ \cite{PRL_renner1998_T} or
TlBa$_2$Ca$_2$Cu$_3$O$_{8.5+\delta}$ \cite{JPSJ_Sekine2016}. As
observed in experiments
\cite{PRL_renner1998_B,PRL_Pan2000,Thesis_Kugler2000}, we find a
similar situation as function of the magnetic field, crossing a
vortex (see Fig.\,\ref{Fig3}, right panel). Here we assume that the
effect of magnetic field is to reduce the coherence energy $\beta$.
Quite analogous to the effect found in case of a disorder potential
\cite{Sci_McElroy2005,PRL_Cren2000}. The coherence peaks decrease
when approaching the vortex center where finally a peakless
pseudogap is found in the vortex core. In a general way, the loss of
coherence of the SC to PG transition is driven by the vanishing of
$\beta$, the pair-pair interaction.

\section{DOS in the pair-pair interaction model}

Contrary to standard models, we express the quasiparticle DOS as a sum of three terms: 
\begin{equation}
\mathcal{N}(E,T)=\mathcal{N}_{cond}(E,T)+\mathcal{N}_{ex}(E,T)+\mathcal{N}_{diss}(E,T)
\label{DOS_sum}
\end{equation}
where the distinct contributions are:
\vspace{0,2cm}

($i$) the condensate, $\mathcal{N}_{cond}(E,T)$

($ii$) the excited pairs, $\mathcal{N}_{ex}(E,T)$

($iii$) the dissociated pairs, $\mathcal{N}_{diss}(E,T)$
\vspace{0,2cm}

\subsection{Condensate}
The contribution of the condensate $\mathcal{N}_{cond}(E,T)$ is proportional to the number of pairs in the condensate at temperature $T$:
\begin{equation}
\mathcal{N}_{cond}(E,T)=N_{oc}(T)\times\mathcal{N}_{SC}^d(E,T)
\end{equation}
where $\mathcal{N}_{SC}^d(E,T)$ is the zero-temperature superconducting DOS, found when all pairs belong to the condensate, i.e. when $N_{oc}(T=0)=n_0$.

As shown in Ref.\,\cite{PRB_sacks2006,SciTech_Sacks2015}, neglecting any QP lifetime broadening, the condensate DOS along the $\theta$ direction is\,:
\begin{eqnarray}
&\mathcal{N}_{SC}^{d}(E, \theta)&=\frac{\mathcal{N}_n(E_F)}{2\,\pi}\ \int_0^\infty d\epsilon_k\
\delta(E_k - E) \hskip 0.1 cm \null \nonumber\\
& \null &=\frac{\mathcal{N}_n(E_F)}{2\,\pi}\
\left[\frac{\partial\epsilon_k}{\partial E_k}\right]_{E_k=E}
\end{eqnarray}
with
$$
\frac{\partial\epsilon_k}{\partial E_k} = \frac{E_k -
\Delta_k(E_k,\theta) \frac{\partial\Delta_k}{\partial
E_k}}{\epsilon_k} = \frac{E_k - \Delta_k(E_k,\theta)
\frac{\partial\Delta_k}{\partial E_k}}{\sqrt{E_k^2 -
\Delta_k(E_k)^2}}
$$
where $\mathcal{N}_n(E_F)$ is the normal DOS at the Fermi energy.
The superconducting DOS is thus inversely proportional to the slope of $E_k(\epsilon_k)$, depending explicitly on the gap function (\ref{gap_equa}). Therefore, the fine structure in the DOS, the `peak-dip-hump' features, originates from the interaction term, $2\beta_k P_0(E_k)$, of the gap equation. While these features have been studied at $T=0$ \cite{PRB_sacks2006,SciTech_Sacks2015}, in the following we address the question of their temperature dependence.

\subsection{Excited pairs}
As soon as the temperature increases, excited pairs start to contribute (see Fig.\,\ref{Fig2}) to the DOS. This leads to the term:
\begin{equation}
\mathcal{N}_{ex}(E,T)=\mathcal{A}(T)\sum_{\varepsilon_i > \delta \cos(2\theta)}f_B(\varepsilon_i,T)\,\mathcal{N}_i(E,\Delta_i)
\end{equation}
where $\varepsilon_i=\Delta_i-\Delta_p\cos(2\theta)$ is the excitation energy and $f_B(E,T)$ is the Bose distribution.
$\mathcal{N}_i(E,\Delta_i)$ is the DOS associated with each pair $\Delta_i$. In the standard {\it d-wave} form, assuming that excited pairs are not interacting, we have:
$$\mathcal{N}_i(E,\Delta_i)=\frac{\mathcal{N}_n(E_F)}{2\pi}\int_0^{2\pi}d\theta\frac{|E|}{\sqrt{E^2-(\Delta_i\cos(2\theta))^2}}$$

\subsection{Dissociated pairs}
\begin{figure}[t]
\centering
\includegraphics[width=8.0 cm]{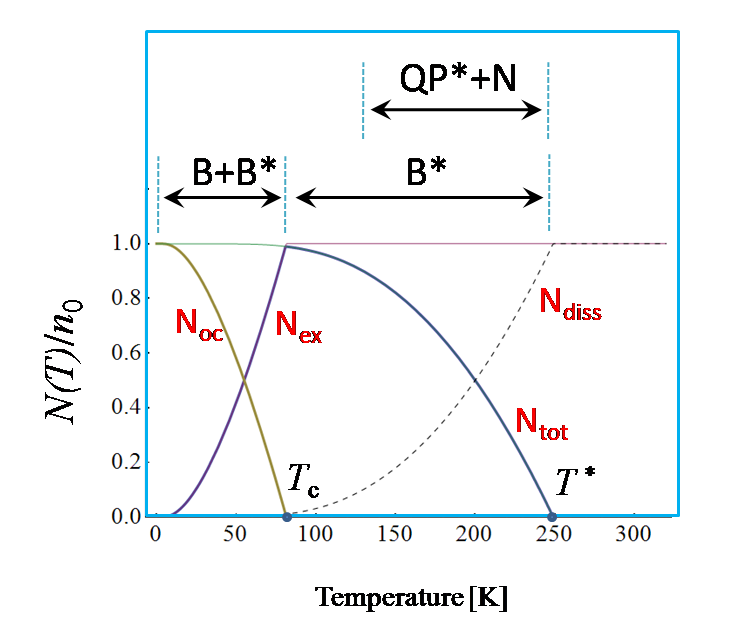} \caption{Plot of the 4
important occupation densities (underdoped case)\,: $n_0$ the total number of pairs at zero temperature, $N_{oc}$ the number of pairs in the condensate, $N_{ex}$ the excited bosons, $N_{diss}$ the dissociated pairs. The upper part of the
figure shows the characteristic temperature ranges in the model\,: B
-- boson of the condensate (vanishing at $T_c$), B* --
excited boson  (maximum at $T_c$ but vanishing at $T^*$), QP*
-- quasiparticles excitations (also vanishing at $T^*$) , N --
`normal' electrons arising from pair dissociation (maximum at $T^*$).}\label{Fig2}
\end{figure}

The bosons in our model are composite fermions but, distinct from
the BCS case, as the temperature rises an increasing fraction of the
pairs {\it dissociate at the Fermi energy}. This appears as some of
the $\Delta_i$ vanish, i.e. when $\Delta_p(T)$ starts to decrease.
This gives rise to the third term:
\begin{equation}
\mathcal{N}_{diss}(E,T)=N_n(E_F)\times N_{diss}(T)
\end{equation}
Here $N_{diss}(T)$ is the number of dissociated pairs at the
temperature $T$ (see Fig.\,\ref{Fig2}). It increases up to $T^*$,
where finally $N_{diss}(T^*)=n_0$.

\section{Discussion}
\subsection{Pair densities}
The number of pairs (including dissociated pairs) must follow a sum rule:
\begin{equation}
N_{oc}(T)+N_{ex}(T)+N_{diss}(T)=n_0
\end{equation}
This rule involving the number of pairs guarantees the conservation
of states in the DOS at any temperature $T$:
$$\int_{-\infty}^{\infty}dE\,\mathcal{N}(E,T)=C$$ where $C$ is a
constant. In addition, we assume that the total number of pairs
(i.e. condensed plus excited pairs) must follow
$$N_{tot}(T)=N_{oc}(T)+N_{ex}(T)\sim\Delta_p(T)^2$$ which vanishes
as $\sim |T-T^*|$ near $T^*$. Note that the precise shape of
$N_{tot}(T)$ has never been investigated. Our assumption on its
particular temperature dependence does not influence the basic
conclusions of the paper.

As previously noted, at $T=0$, all pairs belong to the condensate:
$N_{oc}(T=0)=n_0$. As the temperature increases, pairs are
progressively excited from the condensate, i.e. $N_{ex}(T)$
increases with temperature in opposition to $N_{oc}(T)$, as in
Fig.\,\ref{Fig2}. Only low-lying excited states dominate owing to
the boson occupation $f_B(E,T)$ in equation (\ref{Noc}).  With
temperature, the condensate empties while the number of excited
pairs increase. Finally, at the critical temperature, there is no
longer a condensate:
$$N_{oc}(T_c)=0$$

\begin{figure*}[t]
\centering
\includegraphics[width=16.0 cm]{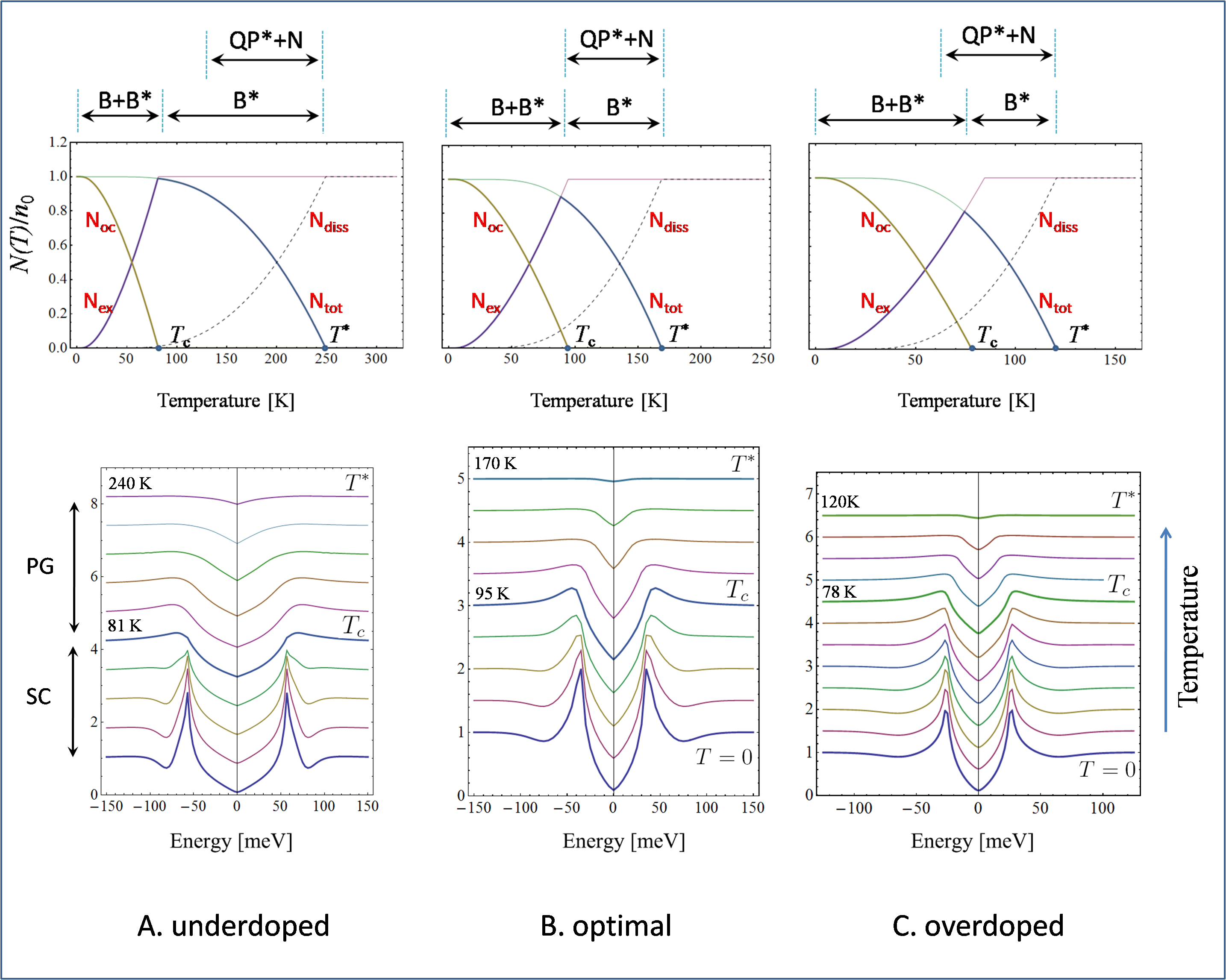} \caption{
Comparison of the $T$-dependent DOS for three different doping
levels\,: underdoped (left panel), optimal (middle panel), overdoped
(right panel). The corresponding temperatures ($T_c$, $T^*$) are
indicated. The upper figures show the temperature range of the boson
and fermion excitations. It is important to note the striking
difference between the onset of quasiparticle and pair dissociation,
for the overdoped and underdoped cases.}\label{Fig4}
\end{figure*}

In the underdoped regime, where $T_c\ll T^*$ (the case of
Fig.\,\ref{Fig2}), the number of dissociated pairs is negligible, so
that all pairs are excited at $T_c$: $$N_{ex}(T_c)\approx n_0$$ In
the more general case, to be discussed below, some pairs are already
dissociated at the critical temperature and one has:
$$N_{diss}(T_c)=n_0-N_{ex}(T_c)>0$$
The effect of pair dissociation is highly sensitive to the
$(T^*-T_c)$ difference, in practice to the doping value. An in-depth
work of Yazdani et
al.\cite{Nat_Gomes2007,PhysicaC_Gomes2007,JphysChem_Gomes2008}
investigated the zero bias conductance in the SC and PG states.
While it is quite uniform in the superconducting state, it showed
significant variations above $T_c$. This result can be direcly
explained by the presence of dissociated pairs, giving rise to the
third term in equation (\ref{DOS_sum}).

It is important to stress that the progressive broadening of the QP
peaks from SC to PG state is not due to a finite quasiparticle
lifetime ($\Gamma_{Dynes}$ broadening \cite{PRL_Dynes1978}).
Instead, it results from the second term in equation\,\ref{DOS_sum}
due to excited pairs existing at finite temperature. The calculation
leads to a larger PG in the vortex core (see Fig.\,\ref{Fig3}, right
panel), where it directly stems from the distribution of preformed
pairs $P_0(\Delta_i)$, rather than at $T_c$ where it also involves
the Bose statistics $f_B(E,T)$. Such a filling of the DOS is a
direct consequence of the statistical occupation of pair excitations
inherent to the model.

\subsection{Changes of the DOS with doping}

Using the previous relations, we have calculated the temperature
evolution of the DOS for three different doping levels corresponding
to the underdoped regime (Fig.\,\ref{Fig4}A), optimally doped
(Fig.\,\ref{Fig4}B) and overdoped regime (Fig.\,\ref{Fig4}C). At low
temperature, the excitation spectrum exhibits the characteristic
{\it d-wave} shape of HTc cuprates (see \cite{fischer_rev} for a
review) and the `peak-dip' features. Note that it is more pronounced
in the underdoped case, while it is slightly attenuated in the
overdoped case.

The coherence peaks broaden as the temperature rises, as a
consequence of the increasing contribution of incoherent excited
pairs, and eventually vanish at $T_c$, where all bosons have left
the condensate. The contribution of $N_{ex}(T)$ gives rise to a
clear pseudogap at $T_c$ for the three different doping values.

The remarkable difference between underdoped and overdoped samples
emerges as the limit at which quasiparticle QP$^*$ and normal
electrons appear (it is defined as the temperature at which the
total number of pairs $N_{tot}(T)$ decreases significantly, i.e.
when it reaches $\sim 0.9\,n_0$). They originate respectively from
excited pairs $\Delta_i$, which give rise to quasiparticles
$E_k^i=\sqrt{\epsilon_k^2+\Delta_i^2}$, and from dissociated pairs
(with energy $\sim 2\varepsilon_k$). The presence of normal
electrons implies that some states are present inside the gap,
leading to a finite value in the DOS at $E=0$. As a rule of thumb,
pair excitations attenuate the QP peaks while the normal states fill
in the gap.

The situation is quite extreme in the underdoped case, wherein $T_c$
is much smaller than $T^*$ (Fig.\,\ref{Fig4}A, upper panel).
Consequently quasiparticles QP$^*$ and normal electrons arise well
above $T_c$. In this case, the SC to PG transition is dominated by
boson excitations, the fermions contribution being negligible. This
quasi-Bose transition is characterized by the vanishing of the
condensate (B) at $T_c$ leaving only pair excitations (B$^*$) just
above $T_c$. The opposite scenario is found in the overdoped regime
where quasiparticles and normal electrons coexist {\it below} $T_c$.
Finally, in the optimally doped regime, with the highest $T_c$, it
is remarkable that the onset of quasiparticle and normal excitations
(QP$^*$+N, in Fig.\,\ref{Fig4}B) coincides with the critical
temperature.
\\Given their importance, further experiments should be done to verify the density profiles with the predictions of the model.

\section{Conclusion}

In summary, we proposed a comprehensive model for HTc
superconductors based on an underlying disordered state of preformed
pairs. Coherence is achieved by means of the pair-pair interactions
leading to a Bose-type condensation. Contrary to the standard BCS
theory, several ingredients contribute to the DOS at finite
temperature: i) the condensate, ii) the presence of excited pairs,
iii) normal electrons arising from pair dissociation.

As a direct consequence of our model, the first contribution
decreases with temperature, while the two other contributions,
preformed pairs and normal electrons, coexist at finite temperature.
The low temperature DOS exhibits the well-known `peak-dip' structure
- directly related to the pair-pair interaction - which
characterizes superconducting coherence. Due to the presence of
excited pairs, the quasiparticle peaks in the DOS decrease
progressively, concomitant with the pair-pair interaction, as a
function of temperature and a pseudogap is finally found at $T_c$ up
to $T^*$ in the whole doping range. A similar pseudogap is found in
the vortex core, in very good agreement with tunneling experiments.

The evolution of the DOS as a function of temperature reveals a mixed state wherein condensed or excited bosons and normal electrons coexist. 

\end{document}